\documentclass[11pt,reqno]{nature}

\usepackage[parfill]{parskip}  
\usepackage[english]{babel}
\usepackage{amsmath}
\usepackage{amsfonts}
\usepackage{amssymb}
\usepackage{amsthm}
\usepackage[pdftex]{graphicx}
\usepackage{epstopdf}
\usepackage{enumitem}
\usepackage{geometry}
\usepackage{hyperref}
\usepackage{tikz}
\usepackage[all]{xy}
\usepackage{multirow}
\usepackage{anysize}
\usepackage{color}
\usepackage{pdflscape}
\usepackage[displaymath, mathlines]{lineno}

\newtheorem*{th1*}{Theorem 1}
\newtheorem*{th2*}{Theorem 2}
\newtheorem*{corollary*}{Corollary}
\newtheorem*{def1*}{Definition 1}
\newtheorem*{def2*}{Definition 2}
\newtheorem*{lemma*}{Lemma}
\newtheorem*{pro1*}{Proposition 1}
\newtheorem*{pro2*}{Proposition 2}
\newtheorem*{pro3*}{Proposition 3}
\newtheorem*{pro4*}{Proposition 4}

\def\wl{\par \vspace{\baselineskip}}
\DeclareMathOperator{\PGG}{PGG}
\DeclareMathOperator{\NPD}{NPD}
\DeclareMathOperator{\PD}{PD}
\newcommand{\beq}{\begin{equation}}
\newcommand{\eeq}{\end{equation}}
\newcommand{\commentout}[1]{}

\def\cents{\hbox{\,\rm\rlap/c}}

\marginsize{3.2cm}{3.2cm}{3.2cm}{3.2cm}

\title{Group size effect on cooperation in one-shot social dilemmas}

\author{H\'el\`ene Barcelo$^1$ $\&$ Valerio Capraro$^2$}

\begin{document}


\maketitle

$^1$Mathematical Sciences Research Institute, CA 94720, Berkeley, USA. $^2$Center for Mathematics and Computer Science (CWI), 1098 XG, Amsterdam,  The Netherlands. V.Capraro@cwi.nl

\begin{center}
\emph{Forthcoming in Scientific Reports}
\end{center}
\wl

\begin{abstract}
Social dilemmas are central to human society. Depletion of natural resources, climate protection, security of energy supply, and workplace collaborations are all examples of social dilemmas. Since cooperative behaviour in a social dilemma is individually costly, Nash equilibrium predicts that humans should not cooperate. Yet experimental studies show that people do cooperate even in anonymous one-shot interactions. In spite of the large number of participants in many modern social dilemmas, little is known about the effect of group size on cooperation. Does larger group size favour or prevent cooperation? We address this problem both experimentally and theoretically. Experimentally, we find that there is no general answer: it depends on the strategic situation. Specifically, we find that larger groups are more cooperative in the Public Goods game, but less cooperative in the N-person Prisoner's dilemma. Theoretically, we show that this behaviour is not consistent with either the Fehr $\&$ Schmidt model or (a one-parameter version of) the Charness $\&$ Rabin model, but it is consistent with the cooperative equilibrium model introduced by the second author. 
\end{abstract}



Social dilemmas are situations in which selfish interest collides
with collective interest. Every individual has an incentive to
deviate from the common good, but if all subjects acted selfishly
they would all be worse off. Depletion of natural resources,
intergroup conflicts, climate protection, security of basic social
systems, workplace collaborations, and price competition in markets
are just some of the fundamental situations that can be modelled
by means of a social dilemma. Consequently, understanding how and
why cooperation can evolve is of primary importance across all biological and social sciences\cite{H64, Ol,Tr, Ax-Ha, H, Ax, S04, No06, Ca13, Ra-No,Ca-Ha,DH,TN,AMFC,ZM,HRPN,SP10,Perc13,Ca-Ma}.

In modern society, many social dilemmas involve a large number of players: firms trying to sell (approximately) the same product, countries involved in reducing the greenhouse gas emissions, taxpayers, work groups, etc. Yet, little is known about the effect of group size on cooperation. Does group size influence cooperation and, if so, how? 

A classical point of view states that cooperation should be more difficult in larger groups: since the increase in the number of people \emph{inevitably} leads to decreasing individual gains relative to the cost of cooperation, free-riding would be much more pervasive in larger groups: large groups will fail; small groups may succeed\cite{Ol, H}. However, others pointed out that the increase in the number of people does not necessarily lead to lower individual gains: the incentive to cooperate can even increase and lead to a positive relation between the size of the group and the level of collective actions\cite{Ch, OM88}. 

These two possibilities can be captured using two well-known economic games.

\textbf{Public Goods Game (PGG).} $N\geq2$ contributors are endowed with $y$ dollars and must simultaneously decide how much, if any, to contribute to a public pool. The total amount in the pot is then multiplied by a constant and evenly redistributed among all players. So the monetary payoff of player~$i$ is $u_i(x_1,\ldots,x_N)=y-x_i+\gamma(x_1+\ldots+x_N)$, where $x_i$ denotes $i$'s contribution, and the `marginal return' $\gamma$ is assumed to belong to the open interval $\left(\frac1N,1\right)$. 

\textbf{$N$-person Prisoner's Dilemma (NPD).} $N\geq2$ agents have the choice of either cooperate (C) or defect (D). To defect means doing nothing, while to cooperate means paying a cost $c>0$ to generate a benefit $b>c$ that gets shared by all other players. So, if $C_{-i}$ denotes the number of agents other than $i$ who cooperate, then agent $i$'s payoff is $\frac{bC_{-i}}{N-1}-c$ , if $i$ cooperates, and $\frac{bC_{-i}}{N-1}$, if $i$ defects. 

The  individual benefit for full cooperation in the PGG is equal to $\gamma N$ and so it increases linearly with the number of players, while the individual cost for cooperation remains constantly equal to $y$. On the other hand, both the individual benefit for full cooperation, $b$, and the individual cost for cooperation, $c$, remain constant in the NPD as the number of players increases, but, in order to reach the benefit, one needs more people to cooperate. With the above discussion in mind, we should then observe a positive effect of group size on cooperation in the PGG and a negative effect of group size on cooperation in the NPD.

Experimental research has not helped much so far, though it has partially confirmed this prediction. Some experimental results suggest that contributions to the public good are slowly increasing with the number of players\cite{McG, IWW}, while others suggest no group size effect\cite{W} or even a negative effect\cite{NQS}. Overall, the picture that emerges is that group size has a moderate, positive effect on cooperation in PGG: in her meta-analysis, Zelmer\cite{Ze} uses data from 27 PGG experiments, conducted using different parametrisations and procedures, and finds a positive, moderate (at the $10\%$ level) effect of group size on contributions. However, these experimental studies have been conducted on iterated games, where long-term strategies and group structure may have played a role in shaping the results.

The effect of group size on cooperation in the NPD is also unclear, but slightly in the opposite direction than the PGG: some studies report no effect\cite{D77}, and others report a negative effect\cite{K82, G12}. The only studies regarding one-shot games\cite{D77,K82} have the issue that, together with the group size, they also vary either $b$ or $c$. Since these latter parameters are known to influence the rate of cooperation in opposite ways\cite{CJR, CSMN, EZ}, it is then difficult to say which parameter has caused which effect. To the best of our knowledge, only one study\cite{G12} keeps $b$ and $c$ constant while varying the size of the group, but it was conducted on iterated games rather than on one-shot games. In this study the authors vary the size of the group from two to five players and find that two people are more cooperative than three, but three people are statistically as cooperative as four or five people.

The experimental contribution of this paper is to clarify this point: Does group size have an effect on cooperation in the  PGG and NPD and, if so, how? Since repetitions may create all sorts of ill-understood noise and spillovers across periods that make it difficult to isolate the effect of the size of a group, we have focused on one-shot games. We have conducted two experiments using the online labour market Amazon Mechanical Turk (AMT)\cite{PCI, HRZ, R12}, one with the PGG and the other one with the NPD. In each of these studies we have separated the participants in two conditions. In Condition~S participants played in small groups and in Condition~L they played the same one-shot social dilemma but in large groups. Our results are in line with the ones discussed above. We report a significant negative effect of group size on cooperation in the NPD and a significant positive effect of group size on cooperation in the PGG.

Our paper also provides a theoretical contribution. Since cooperation is typically enforced by means of external controls, such as punishment for defectors or reputation for cooperators\cite{Ha, Da, MSK, AHV, BGBR, RM, R08, HS}, predicting the expected rate of cooperation without forms of external controls would allow to optimise the use of these techniques, besides shedding light on the cognitive basis of human decision-making. Are there theoretical models predicting our findings? We show that our results are inconsistent with both Fehr $\&$ Schmidt's\cite{Fe-Sc} and (the one parameter version of) Charness $\&$ Rabin's\cite{Ch-Ra} models, but they are consistent with the cooperative equilibrium model\cite{Ca13}, which indeed predicts both regularities observed. Moreover, since the cooperative equilibrium is a parameter-free model, we can make a direct comparison between experimental data and predictions. We show that, though the predictions of the cooperative equilibrium model point qualitatively in the same direction as the experimental data, they are quantitatively still quite off. Further research is needed to understand how the model could be improved to fit the experimental data better even from a quantitative point of view. 

\section*{Experimental results}

We recruited US subjects using the online labour market AMT. As in classical lab experiments, AMT workers receive a baseline payment and can earn an additional bonus depending on how they perform in the game. AMT experiments are easy to implement and cheap to realise, since AMT workers are paid a substantially smaller amount of money than people participating in physical lab experiments. Nevertheless, it has been shown that data gathered using AMT agree both qualitatively and quantitatively with those collected in physical labs\cite{HRZ,R12,SW,RAC}. 

We conducted two studies: in Study A we investigated the group size effect on the Public Goods game and in Study B we investigated the group size effect on the Prisoner's dilemma. For the PGG we opted for small groups of size four and large groups of size forty; for the NPD we opted for small groups of size two and large groups of size eleven. These particular choices are due to the theoretical predictions presented in the next section. Indeed, the cooperative equilibrium model predicts that the size of the group has a weak positive effect on cooperation in the PGG and so here we opted for a small group (four people) and a very large one (forty people); on the other hand, the cooperative equilibrium predicts a strong (exponential) negative effect of the size of he group on cooperation in the NPD. Thus we opted for a very small group $N=2$, because it might be possible that for $N>2$ the rate of cooperation decays already so much as to make every statistical analysis insignificant. The choice $N=11$ for large groups, was made only because we needed a number such that the benefit for cooperation, that is $\$0.30$, could be easily divided by $N-1$.

\emph{Study A. Group size effect on cooperation in the Public Goods Game}

As in other experimental\cite{IWW,W,NQS} and theoretical\cite{SA,SP} works, we study group size effect on cooperation in the Public Goods game by keeping the marginal return for cooperation $\gamma$ fixed and increasing the size of the group. Participants earned $\$0.30$ for participation and were randomly assigned to either of two conditions. In Condition~S (as in small), they were asked to play a four-player PGG with endowment $y=\$0.10$ and constant marginal return $\gamma=0.5$; in Condition~L (as in large), they were asked to play a forty-player PGG with the same endowment and marginal return. After the instructions, participants were asked to answer four comprehension questions. Subjects failing any of them were automatically excluded from the game. Exact full instructions are reported in the Method section.

62 subjects ($31\%$ female, average age $29.1$) passed the comprehension questions in Condition~S and 66 subjects ($51\%$ female, mean age 28.9) passed the comprehension questions in Condition~L. In AMT experiments with comprehension questions it is virtually impossible to stop the experiment when a precise number of participants is reached. In order to compute the payoffs (for instance) in Condition~L, we randomly created two groups of forty people each and we computed the payoffs. Clearly some of the participants belonged to both groups. For these participants we randomly selected either of the two payoffs to be their final bonus. 

Table \ref{ta: public} reports all relevant statistics. SEM denotes the standard error of the mean. It is clear that increasing the group size has the effect that more and more people contribute to the public good. The visual impression is confirmed by the Wilcoxon rank-sum test ($P=0.0002$).

\begin{table}
\centering
\begin{tabular}{| c | c | c | c | c |}
\hline\hline

Condition & $\%$ free-riders & $\%$ contributors & Mean contribution & SEM \\ [0.5ex]
\hline
S & 48.38 &30.64&3.92&0.56\\
L & 21.21&60.60 & 6.91&0.51\\[1ex]
\hline
\end{tabular}
\vspace{0.3cm}
\caption{Descriptive statistics of Study 1. Public Goods Game with 4 players (Condition~S) versus PGG with 40 players (Condition~L). In both conditions, the maximum possible contribution was $\$0.10$ and the marginal return was $\gamma=0.5$. The results show clearly that the larger group is much more cooperative than the smaller group and this is confirmed by the statistical analysis, showing that the means are signifcantly different (Rank-sum, $p=0.0002$).}\label{ta: public}
\end{table}

\emph{Study B. Group size effect on cooperation in the $N$-person Prisoner's Dilemma}

As in our first study, also Study B uses US subjects recruited through the online labour market AMT. Participants earned $\$0.30$ for participation and were randomly assigned to either of two conditions. In Condition~S, they were asked to play a two-player PD with benefit $b=\$0.30$ and cost $c=\$0.10$. in Condition~L, they were asked to play an eleven-player PD with the same benefit and cost. Also here we asked for comprehension questions and we automatically excluded from the game those subjects who fail to correctly answer any of them. Full instructions are reported in the Method section.

75 subjects ($28\%$ female, average age $28.8$) passed the comprehension questions in Condition~S and 78 subjects ($32\%$ female, mean age 29.6) passed the comprehension questions in Condition~L. Table \ref{ta: prisoner} reports all relevant statistics. SEM denotes the standard error of the mean. It is clear that increasing the group size has the effect that fewer and fewer people cooperate. The visual impression is confirmed by Wilcoxon rank-sum test ($P=0.0404$).

\section*{Theoretical results}\label{se:models}

The previous section reported experimental results in support of the following two findings: (i) Group size has a positive effect on cooperation in the Public Goods Game; (ii) Group size has a negative effect on cooperation in the $N$-person Prisoner's Dilemma. Are there mathematical models of human behaviour predicting both these regularities?

Here we first consider two of the most widely used models of human behaviour, the Fehr $\&$ Schmidt\cite{Fe-Sc} and (a one-parameter version of) the Charness $\&$ Rabin's\cite{Ch-Ra} models, and we show that none of them predict the above regularities. Then we extend the cooperative equilibrium model\cite{Ca13} from two-person social dilemmas to \emph{some} social dilemmas and we show that it in fact predicts both of them. Proofs of the results are postponed to the Supplementary Information.

From now on, we fix the following notation. Given a game $\mathcal G$, let $P$ denote the set of players, each of which has pure strategy set $S_i$ and monetary payoff function $u_i$.

\emph{Fehr $\&$ Shmidt model}

Fehr $\&$ Schmidt model\cite{Fe-Sc} assumes that, given the strategy profile $\sigma$, the utility of player~$i$ is
\begin{linenomath*}
\begin{align}
U_i(\sigma)=u_i(\sigma)-\frac{\alpha_i}{N-1}\sum_{j\neq i}\max(u_j(\sigma)-u_i(\sigma),0)-\frac{\beta_i}{N-1}\sum_{j\neq i}\max(u_i(\sigma)-u_j(\sigma),0),
\end{align}
\end{linenomath*}
where $0\leq\beta_i\leq\alpha_i$ are individual parameters. Specifically, $\alpha_i$ represents the extent to which player $i$ is averse to inequity in the favour of others, and $\beta_i$  represents the extent to which player~$i$ is averse to inequity in his favour.

Given a population $\mathcal P$ of players, each of which with parameters $(\alpha_i,\beta_i)$, we denote by $\mu(\mathcal P, \NPD(b,c))$ the percentage of people $i\in \mathcal P$ such that $U_i(C,\ldots, C,D,C,\ldots, C) \leq U_i(C,\ldots, C)$ in the NPD.
\wl
\begin{pro1*}
For fixed $b$ and $c$, the function $\mu(\mathcal P, \NPD(b,c))$ is decreasing with~$N$.
\end{pro1*}

Similarly, consider the PGG with total endowment normalised to be $y=1$, so as full cooperation means to contribute 1 and defection corresponds to contribute 0. Let 
$\mu(\mathcal P, \PGG(\gamma,N))$ denote the percentage of people $i\in\mathcal P$ such that {$U_i(1,\ldots, 1,0,1,\ldots, 1) \leq U_i(1,\ldots,1)$}.
\wl
\begin{pro2*}
For fixed $\gamma$,  the function $\mu(\mathcal P, \PGG(\gamma,N))$ is independent of $N$.
\end{pro2*}

In conclusion, while the results of our Study B are consistent with the Fehr-Schmidt model, those of our Study A are not.

\emph{Charness $\&$ Rabin's model}

We consider the following simple form of the Charness $\&$ Rabin model\cite{Ch-Ra}. Given a strategy profile $\sigma$, we assume that player~$i$ experiences a tension between self interest and common interest and so he or she has utility
\begin{linenomath*}
\begin{align}
U_i(\sigma)=\alpha_i u_i(\sigma)+(1-\alpha_i)\sum_{j=1}^Nu_j(\sigma),
\end{align}
\end{linenomath*}
where $\alpha_i\in[0,1]$ is an individual parameter describing how much player~$i$ cares about the total welfare.

As before, we consider a population $\mathcal P$ and we denote $\mu(\mathcal P, \NPD(b,c))$ the percentage of people $i\in P$ such that $U_i(C,\ldots, C,D,C,\ldots, C) \leq U_i(D,\ldots, D)$ in the NPD.
\wl
\begin{pro3*}
For fixed $b$ and $c$ the function $\mu(\mathcal P, \NPD(b,c))$ is independent of $N$.
\end{pro3*}
\wl
\begin{pro4*}
For fixed $\gamma$, the function $\mu(\PGG(\gamma,N))$ is increasing with $N$.
\end{pro4*}

In sum, this one-parametric version of the Charness $\&$ Rabin model makes qualitatively the right prediction in the PGG, but not in the NPD.

We mention that the most general form of the Charness $\&$ Rabin model takes also into account inequity aversion and uses the utility function
\begin{linenomath*}
\begin{align}
V_i(x_1,\ldots,x_N)=(1-\alpha_i)x_i+\alpha_i(\delta_i\min(x_1,\ldots,x_n)+(1-\delta_i)(x_1+\ldots+x_N)).
\end{align}
\end{linenomath*}
Not surprisingly, this general version makes the right prediction in both the PGG and the NPD (we leave the tedious computation to the reader). What we find surprising, however, is that the ultimate reason for the different predictions is caused by two different effects: the tendency to maximise group welfare favours cooperation in large PGG and inequity aversion prevent cooperation in the NPD. 

We now pass to the description of a model which predicts both regularities by appealing to the same effect (tendency to maximise the group welfare) and without using any free parameter. The price to pay is a major technical difficulty and the fact that the model, at the moment, is definable only for what we call \emph{highly symmetric} games.

\emph{Cooperative equilibrium}

The cooperative equilibrium is a new parameter-free solution concept for two player social dilemmas, which has been shown to make accurate predictions of average behaviour in one-shot\cite{Ca13,Ca13b} and iterated\cite{CVPJ} social dilemmas. It is typically a mixed strategy depending on the payoffs, which organises all classical regularities: The level of cooperation in the Prisoner's Dilemma increases as the cost/benefit ratio decreases; The level of cooperation in the Traveler's dilemma increases as the bonus/penalty decreases; The level of cooperation in the Public Goods Game increases as the constant marginal return increases. We now extend this model to \emph{some} N-player social dilemmas and show that it predicts also the regularities reported in the Experimental Results section.

The key idea behind the cooperative equilibrium is the assumption that humans do not act a priori as single agents, but they forecast how the game would be played if they formed coalitions and then act so as to maximise their forecast. 

A general theory for every normal form game would require to consider all possible coalition structures and deal with the fact that different players may have different forecasts about the same coalition structure. Here we eliminate this problem by restricting to only \emph{highly symmetric} social dilemmas.

\begin{itemize}
\item \textbf{Symmetry.} All players have the same set of strategies $S$ and for each player~$i$, for each permutation $\pi$ of the set of players, and for each strategy profile $(s_1,\ldots,s_N)\in S^N$ one has 
\begin{linenomath*}
\begin{align}
u_i(s_1,\ldots,s_N)=u_{\pi(i)}(s_{\pi(1)},\ldots,s_{\pi(N)}). 
\end{align}
\end{linenomath*}
\item \textbf{High symmetry.} The game is symmetric and has a unique (and symmetric) Nash equilibrium and a unique (and symmetric) profile of strategy maximizing the total welfare. 
\end{itemize}

Given the high level of symmetry of the games under consideration, we consider only the two extremal scenarios: the selfish coalition structure and the fully cooperative coalition structure. 

Let $P$ denote the set of players, each of which has pure strategy set $S$, mixed strategy set $\mathcal P(S)$, and utility function $u_i$. Coalition structures are just partitions of the player set. We denote $p_s$ the selfish coalition structure and $p_c$ the fully cooperative coalition structure. Every coalition structure $p\in\{p_s,p_c\}$ gives rise to a new game $\mathcal G_p$, where players in the same coalition play as a single player aiming to maximise the sum of the payoffs of the players belonging to that coalition. For every highly symmetric social dilemma, $\mathcal G_p$ has a unique Nash equilibrium, which we denote $\sigma^p$. Fix $i\in P$ and let $j\in P\setminus\{i\}$ be another player. We denote $I_{j}(p)$ the maximum payoff that player~$j$ can obtain by leaving the coalition structure $p$. Formally, 
\begin{linenomath*}
\begin{align}\label{eq:incentive}
I_{j}(p):=\max\{u_{j}(\sigma_{-j}^p,\sigma_{j})-u_{j}(\sigma_{-j}^p,\sigma_{j}^p) : \sigma_{j}\in\mathcal P(S)\}.
\end{align}
\end{linenomath*}
$I_{j}(p)$ will be called the \emph{incentive} of player~$j$ to abandon the coalition structure $p$. 

Given a profile of strategies $(\sigma_1,\ldots,\sigma_N)$, a strategy $\sigma_i'\in\mathcal P(S)$ is called an \emph{$i$-deviation} from $(\sigma_1,\ldots,\sigma_N)$ if $u_i(\sigma_{i}',\sigma_{-i})\geq u_i(\sigma_1,\ldots,\sigma_N)$.  

We denote $D_{j}(p)$ the maximal loss that players~$j$ can incur if he decides to leave the coalition structure $p$ to try to achieve his maximal possible gain, but also other players deviate from the coalition structure $p$ to either follow their selfish interests or anticipate player~$j$'s deviation. Formally, 
\begin{linenomath*}
\begin{align}\label{eq:risk}
D_{j}(p):=\max\{u_{j}(\sigma_i^p,\sigma_{-i}^p)-u_{j}(\sigma_j,\sigma_{-j})\},
\end{align}
\end{linenomath*}
where $\sigma_{j}$ runs over the set of strategies such that $u_{j}(\sigma_{-j}^p,\sigma_{j})$ is maximized and $\sigma_{-j}$ runs over the set of profiles of strategies $(\sigma_k)_{k\neq j}$ for which there is $h$ such that $\sigma_h$ is an $h$-deviation from either $\sigma^p$ or $(\sigma^p_{j},\sigma_{-j})$. $D_{j}(p)$ is called the \emph{disincentive} for player~$j$ in abandoning the coalition structure $p$. The number 
\begin{linenomath*}
\begin{align*}
\tau_{i,j}(p):=\frac{I_{j}(p)}{I_{j}(p)+D_{j}(p)}
\end{align*}
\end{linenomath*}
will be informally interpreted as the probability that player~$i$ assigns to the event ``\emph{player~$j$, knowing that all other players are thinking about playing according to $p$, abandons the coalition structure $p$}''. In the context of anonymous games, where the reasoning of a player cannot affect the reasoning of another player, we define, for $J\neq\emptyset$,
\begin{linenomath*}
\begin{align*}
\tau_{i,J}(p):=\prod_{j\in J}\tau_{i,j}(p).
\end{align*}
\end{linenomath*}
Now to define $\tau_{i,\emptyset}(p)$, which is the probability that nobody abandons the coalition structure, we use the law of total probabilities. Assume, for simplicity, that $\tau_{i,j}(p)$ does not depend on $i$ and $j$, as is the case in symmetric games. We find
\begin{linenomath*}
\begin{align*}
\tau_{i,\emptyset}(p)
&=1-\sum_{k=1}^{N-1}(-1)^{k+1}\binom{N-1}{k}\tau_{i,j}(p)^k\\
&=1+\sum_{k=1}^{N-1}(-1)^{k}\binom{N-1}{k}\tau_{i,j}(p)^k\\
&=\sum_{k=0}^{N-1}(-1)^{k}\binom{N-1}{k}\tau_{i,j}(p)^k\\
&=\sum_{k=0}^{N-1}\binom{N-1}{k}(-\tau_{i,j}(p))^k.\\
\end{align*}
\end{linenomath*}
Using Newton's binomial law we finally find
\begin{linenomath*}
\begin{align*}
\tau_{i,\emptyset}(p)=(1-\tau_{i,j}(p))^{N-1}.
\end{align*}
\end{linenomath*}
Now, let $e_{i,\emptyset}(p)$ be the minimum payoff for player~$i$ if nobody abandons the coalition structure $p$, which is just $u_i(\sigma^p)$, and, finally, let $e_{i,J}(p)$ be the infimum of payoffs of player~$i$ when she plays $\sigma_i^p$ but at least one player~$j\in J\setminus\{i\}$ plays a $j$-deviation from $\sigma^p$. The \emph{forecast} of player~$i$ associated to the coalition structure $p$ is defined as
\begin{linenomath*}
\begin{align}
v_i(p):=\sum_{J\subseteq P\setminus\{i\}}e_{i,J}(p)\tau_{i,J}(p).
\end{align} 
\end{linenomath*}
Observe that symmetry implies that the forecast $v_i(p)$, if $p\in\{p_s,p_c\}$, actually does not depend on $i$ and so there is a coalition structure $\overline p$ (independent of $i$) which maximises the forecast for all players. Moreover, $v_i(\overline p)=v_j(\overline p)$, for all $i,j\in P$. We denote this number $v(\overline p)$ and we use it to define common beliefs or, in other words, to make a tacit binding among the players. 
\wl
\begin{def1*}
{\rm The induced game $\text{Ind}(\mathcal G,\overline p)$ is the same game as $\mathcal G$ except for the set of profiles of strategies: the induced game contains only those strategy profiles $\sigma$ such that $u_i(\sigma)\geq v(\overline p)$, for all $i\in P$. }
\end{def1*}

The induced game does not depend on the maximizing coalition structure, that is, in case of multiple coalition structures maximising the forecast, one can choose one of them casually to define the induced game and this game does not depend on such a choice. 

Since the set of strategy profiles in the induced game is convex and compact (and non-empty) one can compute Nash equilibria of the induced game.
\wl
\begin{def2*}
{\rm A cooperative equilibrium for $\mathcal G$ is a Nash equilibrium of the game $\text{Ind}(\mathcal G,\overline p)$.}
\end{def2*}

It has been proven in \cite{Ca13} that this model organises all classical observations on two-person social dilemmas: The rate of cooperation in the Prisoner's Dilemma increases as the cost/benefit ratio decreases; the rate of cooperation in the Traveler's dilemma increases as the bonus/penalty decreases; the rate of cooperation in the Public Goods Game increases as the constant marginal return increases. We now show that it organises also the regularities observed in our experiments.

Let PGG$(N,\gamma)$ denote the Public Goods Game with $N$ players and constant marginal return $\gamma$. To simplify the formulas, we assume that the total endowment of each player is normalized to $y=1$. Denote
\begin{linenomath*}
\begin{align}
v(\gamma,N)=\gamma N\left(\frac{\gamma N-1}{\gamma(N-1)}\right)^{N-1}+\gamma\left(1-\left(\frac{\gamma N-1}{\gamma(N-1)}\right)^{N-1}\right).
\end{align}
\end{linenomath*}
\begin{th1*}\label{th:pgg}
{\rm The only (pure) cooperative equilibrium of the PGG$(N,\gamma)$ is to contribute
\begin{linenomath*}
\begin{align}
\max\left(0,\frac{v(\gamma,N)-1}{\gamma N-1}\right).
\end{align}
\end{linenomath*}
In particular, it is increasing with the variable $N$.}
\end{th1*}

Let $\PD(b,c,N)$ denote the $N$-person Prisoner's Dilemma with cost $c$ and benefit $b$. Denote
\begin{linenomath*}
$$
v(b,c,N)=(b-c)\left(1-\frac{c}{b}\right)^{N-1}-c\left(1-\left(1-\frac{c}{b}\right)^{N-1}\right).
$$
\end{linenomath*}
\wl
\begin{th2*}\label{th:prisoner}
{\rm The only cooperative equilibrium of $\PD(b,c,N)$ predicts cooperation with probability 
\begin{linenomath*}
$$ \lambda=\max\left(0,\frac{v(b,c,N)}{b-c}\right).$$ 
\end{linenomath*}
So cooperation is predicted to decrease with the number of agents.}
\end{th2*}

\emph{Comparison between predictions and experimental data}

Table \ref{ta:model comparison} summarizes the comparison between the models in consideration. Fehr \& Schmidt's model makes qualitatively the right prediction in the NPD but not in the PGG; the one-parameter version of the Charness \& Rabin model makes qualitatively the right prediction in the PGG but not in the NPD. The two-parameter version of the Charness \& Rabin model makes qualitatively the right prediction for both games, at the price of using two free parameters. The cooperative equilibrium makes qualitatively the right prediction for both games, without using any free parameter. Since the cooperative equilibrium is parameter-free, we can make a direct comparison between its predictions and experimental data. Results are summarised in Table \ref{ta:public2} and Table \ref{ta:prisoner2}. While the qualitative behaviour is well-captured, the quantitative prediction is still quite off. An interesting topic for further research is to understand what can be done to improve the quantitative predictions.

\section*{Discussion}\label{se:conclusion}

We have studied how the size of a group influences cooperation in two one-shot social dilemmas, the Public Goods Game and the $N$-person Prisoner's Dilemma. The reason why we have considered these two games is that we expected opposite results. In the PGG the invidual benefit for full cooperation is equal to $\gamma N$ and so it increases linearly with the size of the group, while the cost of cooperation remains constantly equal to $y$; in the NPD both the individual benefit for full cooperation $b$ and the cost of cooperation  $c$ remain constant, but, in order to reach the benefit, one needs more people to cooperate. This difference suggests that we should see a positive effect of group size on cooperation in the PGG and a negative effect of group size on cooperation in the NPD.

To test this prediction we have conducted two experiments using the online labour market Amazon Mechanical Turk. Our results confirmed it showing that forty players are significantly more efficient in providing the public good than only four players (P=0.0002), and that two players are significantly more cooperative than eleven in a Prisoner's Dilemma (P=0.0404). 

Although the positive effect of group size on cooperation in the PGG has long been debated\cite{Ze,L}, we do not find it to be very surprising. It is indeed predicted by the Charness $\&$ Rabin model. Also theoretical research on \emph{iterated} spatial PGGs shows that group size should have a positive effect on cooperation, at least when the PGG is played on a lattice and, after each round, a vertex can take one of his neighbours' strategy with probability depending on the payoff difference\cite{SP}.

Yet, surprisingly, we have found that neither the Fehr $\&$ Schmidt model nor (a one-parameter version of) the Charness $\&$ Rabin model predict both the regularities observed. The general form of Charness $\&$ Rabin's utility makes the right prediction, at the price of using two free parameters. Moreover, different predictions are due to different causes: tendency to maximise the group welfare favours cooperation in large PGGs and inequity aversion prevents cooperation in large NPDs.

We have then extended the cooperative equilibrium model from two-player social dilemmas to \emph{some} N-player symmetric games, that we have called \emph{highly symmetric}. Despite the fact that this model does not use any free parameter, it is able to predict both the above mentioned regularities by appealing to a single cause: tendency to maximise total welfare. The quantitative predictions of the model are not dramatically far from the data we gathered, but, at least in three out of four cases, they are still quite off. More experimental data are needed to understand if and how the model can be improved to fit the data better also from a quantitative point of view. In a one-shot setting, as the one under consideration, it is likely that a substantial proportion of players make computational mistakes when they reason about the strategy they want to choose. Thus, a promising direction of research is to define a sort of quantal cooperative equilibrium in a similar way as the standard notion of quantal response equilibrium\cite{quantal} is defined. Ideally, the quantal cooperative equilibrium would be a one-parameter solution concept where the parameter represents the extent to which the players make mistakes in the computations needed to compute the cooperative equilibrium. 

Other research questions are worth being mentioned. One way to look at our experimental results is as follows. By leaving both the cost and the benefit of cooperation constant, the group size has a negative effect on cooperation, while, when the increase in the number of agents correspond to an increase in the benefit of cooperation (or, equivalently, to a decrease in the cost of cooperation), then the size of the group \emph{may} have a positive effect on cooperation which is ultimately due to the increase of the benefit of cooperation (or to the decrease of the cost of cooperation). Our results thus suggest that the \emph{pure} effect of the group size on cooperation is negative: more people are less likely to cooperate. The group size might have a positive effect only when it generates an increase of the benefit of cooperation or a decrease of the cost of cooperation. Since the difference between the benefit for cooperation and the cost of cooperation in the Public Goods game can be seen as an externally imposed reward to enforce cooperation, it is interesting to figure out what is the \emph{cheapest} sequence (parametrized by the size of the group) of rewards that gives rise to a positive effect of the size of the group on cooperation. More precisely, let $1<b_N<N$ be a non-decreasing sequence of real number. Define the $N$-player \emph{general} Public Goods game to be the $N$-player game where each player can either contribute $0$ or $1$ and gets monetary payoff
\begin{linenomath*}
$$u_i(x_1,\ldots,x_N)=1-x_i+\frac{b_N}{N}\sum_{j=1}^Nx_j,$$
\end{linenomath*}
where $x_j$ denotes player $j$'s contribution. Our results show that the group size has a positive effect on cooperation when $b_N$ increases linearly with the number of players ($b_N=\gamma N$ in the standard PGG) and it has a negative effect on cooperation when $b_N$ is constant in $N$ (which is essentially equivalent to the NPD). Consequently, there must be a sequence $b_N$ of numbers representing the behavioural transition from a negative effect of the group size on cooperation to a positive effect. An interesting question would be to find this sequence.

Our model also predicts existence of strategic situations for which group size has a non-linear effect with intermediate groups being more cooperative. Theorems 1 and 2 indeed imply that a general PGG defined by a sequence $b_N=\gamma N$, for $N\leq N_0$, and $b_N=b_{N_0}$, for all $N>N_0$, will have a cooperative equilibrium whose maximum level of cooperation is reached for some intermediate group size. This strategic situation is likely to happen in real-life situations, in which when the number of cooperators reaches a saturation threshold, no additional benefit is created. A concrete example of such a situation has been recently reported in ref. 53. Here the authors report a field study in which intermediate groups were more cooperative than small and large groups. The case reported is that of monitoring the Wolong Nature Reserve, Sichuan Province, China, by means of a number of households. They found that if the number of households is either too small or too large, then the monitoring effort of each household is also small; the monitoring effort of the households is maximized when the number of households is somewhat intermediate.

Other research questions come from looking at some of the predictions made by the cooperative equilibrium model.
\begin{itemize}
\item For fixed $N$ and $c$, the benefit $b$ has a positive effect on cooperation in the NPD.
\item For fixed $N$ and $b$, the cost $c$ has a negative effect on cooperation in the NPD.
\end{itemize}

Confirmation of these two predictions have been recently found in the case of two players\cite{CJR, CSMN, EZ} and we cannot find any reasonable motivation why they should fail in larger groups.

The model also predicts that, with $N$ fixed, $\gamma$ has a positive effect on cooperation in the PGG. This fact has been confirmed by several experimental studies on both one-shot and iterated PGGs\cite{Ze, IWT, L, SN, Gu}. 

Another highly symmetric social dilemma to which the cooperative equilibrium model can be applied is the Bertrand Competition (BC). In the BC, $N\geq2$ firms compete to sell their identical product. Each of the firms can choose a price between the `price floor' $L$ and the `reservation value' $H>L$. The firm that chooses the lowest price, say $s$, sells the product at that price, getting a payoff of $s$; all other firms get nothing. Ties get split among all firms that made the corresponding price. One easily sees that the only cooperative equilibrium of this game is to set the price
\begin{linenomath*}
\begin{align}
\max\left(L,H\cdot\left(\frac{H}{(H-1)N}\right)^{N-1}\right).
\end{align}
\end{linenomath*}
So the cooperative equilibrium makes the prediction that the size of the group has a strong (more than exponential) negative effect on cooperation in the BC. This prediction has been partially confirmed by other experimental studies\cite{DG, DG02, Br}, which have shown that four people are already enough to completely destroy cooperative behaviour. However, these studies concern iterated games and so more experimental evidence would be needed. 

On the theoretical side, many questions concerning the model should be addressed, starting from the broad question of extending the cooperative equilibrium model as far as possible, at least to include other relevant social dilemmas such as the Volunteer dilemma\cite{D85} and the collective-risk social dilemma\cite{M08}. Unfortunately, these social dilemmas possess one of the general characteristics preventing the development of a general theory: in case the game $\mathcal G_p$ has many (possibly infinite) equilibria, which one should be used as a reference strategy profiles $\sigma^p$ to compute incentives and disincentives? Other theoretical questions include: Can the cooperative equilibrium be expressed in terms of classical utility theory through a utility function to be maximised? 


\section*{Method}

The experimental part of this article uses US subjects recruited through Amazon Mechanical Turk. Participants were randomly assigned to one of four experiments using economic games (PGGlarge, PGGsmall, NPDlarge, NPDsmall).  After entering the survey, AMT workers were asked to type their Turk ID and a long CAPTCHA-like code. Specifically, in order to filter out lazy participants, we asked them to write the following neutral sentence (taken from Wikipedia) in reverse order.

\emph{Morocco has a coast on the Atlantic Ocean that reaches past the Strait of Gibraltar into the Mediterranean Sea. It is bordered by Spain to the north, Algeria to the east, and Western Sahara to the south. Since Morocco controls most of Western Sahara, its de facto southern boundary is with Mauritania.}

Participants who passed this first test were presented the rules of the game. In Condition PGGsmall they were presented as follows:

\emph{You are part of a group of four participants. The amount of money you can earn depends on each of the four participants' decisions.} 
 
\emph{All participants are given 10\cents\ and each one has to decide how much, if any, to contribute to a pool. Your gain will be what you keep plus half of the total amount contributed by all players.}
 
\emph{So, for instance:}
\begin{itemize}
\item \emph{If everyone contributes all of their money, then you end the game with 20\cents.}
\item \emph{If you do not contribute anything and everyone else contributes all of their money, then you end the game with 25\cents.}
\item \emph{If no one contributes anything, then you end the game with 10\cents.}
\end{itemize} 

Instructions for Condition PGGlarge were exactly the same, a part from obvious changes. Instructions for Condition NPDsmall were instead presented as follows:

\emph{You have been paired with another participant. The amount of money you can earn depends on your and the other participant's decision.} 
 
\emph{You are both given 10\cents\ and each of you must decide whether to keep it or give it away. Each time a participant gives away their 10\cents, the other participant earns 30\cents.}
 
\emph{So:}
\begin{itemize}
\item \emph{If you both decide to give the 10\cents, you end the game with 30\cents.}
\item \emph{If you keep it and the other participant gives it away, you end the game with 40\cents.}
\item \emph{If you give it away and the other participant keeps it, you end the game with 0\cents.} 
\item \emph{If you both keep it, then you end the game with 10\cents.}
\end{itemize}
Instructions for Condition NPDlarge were exactly the same, a part from obvious changes.

In order to have good quality results, after explaining the rules of the game, we asked four comprehension questions. Participants who failed any of the comprehension questions were automatically excluded from the game. We could do this very easily using the Survey builder Qualtrics, which allows us to use skip logics: programs that automatically end the survey if the correct answer is not selected. Questions were formulated in such a way to make clear the duality between maximising one's own payoff and maximising the others' payoff. Specifically, we asked the following questions. 
\begin{enumerate}
\item \emph{What is the choice you should make to maximise your gain?}
\item \emph{What is the choice you should make to maximise the other participants' gains?}
\item \emph{What choice should the other participants make to maximise their own gains?}
\item \emph{What choice should the other participants make to maximise your gain?}
\end{enumerate}

Subjects who passed the comprehension questions were then asked to make their decision. After playing, subjects were asked a few basic demographic questions (gender, age, and level of education) and the reason why they made their decision. After this, the survey ended providing the code to claim for the bonus. Participants were also informed that computation and payment of the bonuses would be made at the end of the experiment. No deception was used. Written consent was obtained by all participants, and the experiments were approved by the Southampton University Ethics Committee on the Use of Human Subjects in Research and carried out in accordance with the approved guidelines. 

\begin{addendum}
  \item[Author Contributions] H.B and V.C. designed and performed the experiment, analysed the data and wrote the manuscript.
 \item[Competing Interests] The authors declare that they have no competing financial interests.
\end{addendum}

\begin{table}
\centering
\begin{tabular}{| c | c | c | c | c |}
\hline\hline

Condition & $\%$ free-riders & $\%$ contributors & Mean contribution & SEM \\ [0.5ex]
\hline
S & 48.38 &30.64&3.92&0.56\\
L & 21.21&60.60 & 6.91&0.51\\[1ex]
\hline
\end{tabular}
\vspace{0.3cm}
\caption{Descriptive statistics of Study 1. Public Goods Game with 4 players (Condition~S) versus PGG with 40 players (Condition~L). In both conditions, the maximum possible contribution was $\$0.10$ and the marginal return was $\gamma=0.5$. The results show clearly that the larger group is much more cooperative than the smaller group and this is confirmed by the statistical analysis, showing that the means are signifcantly different (Rank-sum, $p=0.0002$).}\label{ta: public}
\end{table}

\begin{table}
\centering
\begin{tabular}{| c | c | c |}
\hline\hline
Condition & $\%$ cooperators & SEM \\ [0.5ex]
\hline
S & 41.33 &4.87\\
L & 25.64&4.97\\[1ex]
\hline
\end{tabular}
\vspace{0.3cm}
\caption{Descriptive statistics of Study 2. Prisoner's Dilemma with 2 players (Condition~S) versus Prisoner's Dilemma with 11 players (Condition~L). In both conditions, $b=\$0.30$ and $c=\$0.10$. The results suggest that larger groups are less cooperative than smaller ones. This is confirmed by the statistical analysis, which show that the means are significantly difference (Rank-sum, $p=0.0404$).}\label{ta: prisoner}
\end{table}

\begin{table}
\centering
\begin{tabular}{| c | c | c | c |}
\hline\hline

Model & PGG & NPD & free parameters\\ [0.5ex]
\hline
FS & none &negative&two\\
CR1 & positive&none & one\\
CR2 & positive&negative & two\\
CE&positive&negative&none\\[1ex]
\hline
\end{tabular}
\vspace{0.3cm}
\caption{Summary of the predictions of the models under consideration. FS denotes the Fehr \& Schmidt model, which predicts no group size effect on cooperation in the PGG, a negative effect of group size on cooperation in the NPD, and uses two free parameters; CR1 denotes the Charness \& Rabin model with only one free parameter; CR2 denotes the Charness \& Rabin model with two parameters; finally, CE denotes the cooperative equilibrium model.}\label{ta:model comparison}
\end{table}

\begin{table}
\centering
\begin{tabular}{| c | c | c |}
\hline\hline
Condition & Mean contribution & CE prediction \\ [0.5ex]
\hline
S &3.92&0\\
L & 6.91&3.46\\[1ex]
\hline
\end{tabular}
\vspace{0.3cm}
\caption{Comparison between predictions of the cooperative equilibrium and experimental data in the PGG. In Condition~S, the cooperative equilibrium coincides with the Nash equilibrium. In Condition~L, it predicts that players should contribute about $35\%$ of their endowment. CE correctly predicts a positive effect of group size on cooperation.}\label{ta:public2}
\end{table}

\begin{table}
\centering
\begin{tabular}{| c | c | c |}
\hline\hline
Condition & $\%$ cooperators & CE prediction \\ [0.5ex]
\hline
S & 41.33 &50\\
L & 25.64&0\\[1ex]
\hline
\end{tabular}
\vspace{0.3cm}
\caption{Comparison between the prediction of the cooperative equilibrium and experimental data. The cooperative equilibrium coincides with the Nash equilibrium in Condition~L and predicts that half of the people cooperate in Condition~S. The cooperative equilibrium correctly predicts a negative effect of group size on cooperation in the NPD.}\label{ta:prisoner2}
\end{table}

\pagebreak

\begin{huge}
\begin{center}
Supplementary Information
\end{center}
\end{huge}
In this Supplementary Information we collect the proofs of all the results we stated in the Theoretical Results section.

\begin{proof}[Proof of Proposition 1.]
Fix, for notational simplicity, $i=1$. We have $U_1(C,\ldots,C)=b-c$ and
\begin{linenomath*}
\begin{align*}
U_1(D,C,\ldots,C)
&= b-\beta_1\left(b-b\frac{N-2}{N-1}+c\right)\\
&=b-\frac{b\beta_1}{(N-1)}-\frac{c\beta_1}{N-1}.
\end{align*}
\end{linenomath*}
It is clear that if $N$ is large enough (depending on the player and on $b$ and $c$), one has $U_1(D,C,\ldots,C)>U_1(C,\ldots,C)$.
\end{proof}

\begin{proof}[Proof of Proposition 2.]
Fix player~$i=1$. We have $U_1(1,\ldots,1)=\gamma N$ and
\begin{linenomath*}
\begin{align*}
U_1(0,1,\ldots,1)
&= \gamma(N-1)+1-\beta_1\left(\gamma(N-1)+1-\gamma(N-1)\right)\\
&=\gamma(N-1)+1-\beta_1.
\end{align*}
\end{linenomath*}
It is clear then clear that the condition $U_1(0,1,\ldots,1)>U_1(1,\ldots,1)$ is independent of $N$.
\end{proof}

\begin{proof}[Proof of Proposition 3.]
Fix $i=1$. We have $U_1(C,\ldots,C)=\alpha_1(b-c)+(1-\alpha_1)(b-c)(N-1)$ and
\begin{linenomath*}
\begin{align*}
U_1(D,C,\ldots,C)
&=\alpha_1b +(1-\alpha_1)b + (1-\alpha_1)(N-1)\left(\frac{b(N-2)}{N-1}-c\right)\\
&=b+(1-\alpha_1)(bN-2b-cN+c).
\end{align*} 
\end{linenomath*}
Observe that the condition $U_1(D,C,\ldots C)>U_1(C,\ldots,C)$ reduces to $\alpha>1-\frac{c}{b}$ and so it does not depend on $N$.
\end{proof}

\begin{proof}[Proof of Proposition 4.]
We have $U_1(1,\ldots,1)=\alpha_1\gamma N+\gamma N^2-\alpha_1\gamma N^2$ and
\begin{linenomath*}
\begin{align*}
U_1(0,1,\ldots,1)
&=\alpha_1(1+\gamma(N-1))+(1-\alpha_1)(1+\gamma(N-1))+\gamma(1-\alpha_1)(N-1)^2\\
&=1-\gamma N+\gamma N^2+2\alpha_1\gamma N-\alpha_1\gamma N^2-\alpha_1\gamma.
\end{align*}
\end{linenomath*}
It is then clear that the condition $U_1(0,1,\ldots,1)<U_1(1,\ldots,1)$ reduces to 
\begin{linenomath*}
$$
1-\alpha_1\gamma + \gamma N(\alpha_1-1)<0,
$$
\end{linenomath*}
which is always verified if $N$ is large enough.
\end{proof}

\begin{proof}[Proof of Theorem 1.]
Since $\mathcal G_{p_s}=\mathcal G$, then $\sigma^{p_s}$ is the Nash equilibrium of the original game. Since there is no incentive to deviate from a Nash equilibrium, the $\tau$ measure is the Dirac measure concentrated on $J=\emptyset$. Therefore $v_i(p_s)$ coincides with the payoff in equilibrium; that is, $v_i(p_s)=1$.

Let now $p_c$ be the fully cooperative coalition structure and observe that, for all $j\in P$, one has $I_j(p_c)=1-\gamma$ and $D_j(p_c)=\gamma N-1$. Consequently, $\tau_{i,j}(p_c)=\frac{1-\gamma}{\gamma(N-1)}$, for all $i,j\in P$, $i\neq j$. Now, $e_{i,J}(p_c)=\gamma$, for all $J\neq\emptyset$, and $e_{i,\emptyset}(p_c)=\gamma N$. Therefore,
\begin{linenomath*}
\begin{align*}
v_i(p_c)
&=\gamma N\left(1-\frac{1-\gamma}{\gamma(N-1)}\right)^{N-1}+\gamma\left(1-\left(1-\frac{1-\gamma}{\gamma(N-1)}\right)^{N-1}\right)\\
&=\gamma N\left(\frac{\gamma N-1}{\gamma(N-1)}\right)^{N-1}+\gamma\left(1-\left(\frac{\gamma N-1}{\gamma(N-1)}\right)^{N-1}\right).
\end{align*}
\end{linenomath*}
To compute the cooperative equilibrium, we observe that this would be the lowest contribution among the ones which, if contributed by all players, would give to all players a payoff of at least $v_i(p_c)$. To compute this contribution it is enough to solve the equation
\begin{linenomath*}
\begin{align*}
1-\lambda +\gamma N\lambda=v_i(p_c),
\end{align*}
\end{linenomath*}
whose solution is indeed $\lambda=\frac{v_i(p_c)-1}{\gamma N-1}$, as stated. 

It remains to show that the cooperative equilibrium is increasing with $N$. To this end, we replace $N$ by a continuous variable $x\geq2$ and denote $v(x):=v_i(p_c)$, $f(x)=\frac{v(x)-1}{\gamma x-1}$, and $r(x)=\left(\frac{\gamma x-1}{\gamma(x-1)}\right)^{x-1}$. Observe that all these functions are differentiable in our domain of interest $x\geq2$. Our aim is to show that $f(x)$ is increasing, that is, $f'(x)>0$. We start by observing that $r(x)$ is increasing. This can be seen essentially in the same way as one sees the standard fact that $\left(1+\frac{1}{n}\right)^n$ is increasing in $n$, by using Bernoulli's inequality. Hence, we have 
\begin{linenomath*}
$$v'(x)=\gamma r(x)+\gamma r'(x)(x-1)>\gamma r(x).$$
\end{linenomath*} 
Consequently, using also the fact that $\gamma xr(x)=v(x)-\gamma(1-r(x))$, we conclude 
\begin{linenomath*}
\begin{align*}
f'(x)
&=\frac{v'(x)(\gamma x-1)-\gamma(v(x)-1)}{(\gamma x-1)^2}\\
&>\frac{\gamma(\gamma xr(x)-v(x)+1)}{(\gamma x-1)^2}\\
&=\frac{\gamma(1-\gamma(1-r(x)))}{(\gamma x-1)^2}\\
&>0,
\end{align*}
\end{linenomath*}
where, the last inequality follows from the fact that both $\gamma$ and $r(x)$ are strictly smaller than 1.
\end{proof}

\begin{proof}
The forecast associated to the selfish coalition structure is $v_i(p_s)=0$, for all players, corresponding to the payoff in (Nash) equilibrium. To compute the forecast associated to  the fully cooperative coalition structure, observe that $e_{i,\emptyset}(p_c)=b-c$, corresponding to Pareto optimum where all players cooperate. The incentive to deviate from the cooperative strategy is $I_j(p_c)=c$, while the disincentive is $D_j(p_c)=b-c$, corresponding to the loss incurred in case all other players anticipate player j's defection and decide to defect as well. Finally, $e_{i,J}(p_c)=-c$, for all $J\neq\emptyset$, corresponding to the strategy profile where only player~$i$ cooperates and all other players defect. Hence we have 
\begin{linenomath*}
$$v_i(p_c)=(b-c)\left(1-\frac{c}{b}\right)^{N-1}-c\left(1-\left(1-\frac{c}{b}\right)^{N-1}\right).$$
\end{linenomath*}
Of course, if $v_i(p_c)\leq0$, then the cooperative equilibrium coincides with the Nash equilibrium. Otherwise, by symmetry, it is the only strategy $\sigma$ such that
\begin{linenomath*} 
\begin{align}\label{eq:PDsolution}
u_i(\sigma,\ldots,\sigma)=v_i(p_c),
\end{align}
\end{linenomath*}
for all $i\in P$. Setting $\sigma=\lambda C+(1-\lambda)D$, we obtain
\begin{linenomath*}
\begin{align*}
u_i(\sigma,\ldots,\sigma)
&=\lambda\sum_{k=0}^{N-1}\lambda^{N-1-k}(1-\lambda)^k\binom{N-1}{k}\left(\frac{b(N-1-k)}{N-1}-c\right) \\
&\quad+(1-\lambda)\sum_{k=0}^{N-1}\lambda^{N-1-k}(1-\lambda)^k\binom{N-1}{k}\left(\frac{b(N-1-k)}{N-1}\right)\\
&=\sum_{k=0}^{N-1}\lambda^{N-1-k}(1-\lambda)^k\binom{N-1}{k}\frac{b(N-1-k)}{N-1}-c\lambda\\
&=b-c\lambda-\frac{b}{N-1}\sum_{k=0}^{N-1}\lambda^{N-1-k}(1-\lambda)^k\binom{N-1}{k}k.
\end{align*}
\end{linenomath*}
Now we use the fact that
\begin{linenomath*}
\begin{align*}
\sum_{k=0}^{N-1}\lambda^{N-1-k}(1-\lambda)^k\binom{N-1}{k}k=(1-\lambda)(N-1),
\end{align*}
\end{linenomath*}
to reduce Equation (\ref{eq:PDsolution}) to
\begin{linenomath*}
\begin{align}
\lambda(b-c)=v_i(p_c),
\end{align}
\end{linenomath*}
which concludes the proof.
\end{proof}

\end{document}